# Real-Time Prediction of Lower Limb Joint Kinematics, Kinetics, and Ground Reaction Force using Wearable Sensors and Machine Learning

Josée Mallah, Yu Zhu, Kailang Xu, Gurvinder S. Virk, *Senior Member, IEEE,* Shaoping Bai, *Senior Member, IEEE,* and Luigi G. Occhipinti*, *Senior Member, IEEE*

***Abstract*— *Objective:* Walking is a key movement of interest in biomechanics, yet gold-standard data collection methods are time- and cost-expensive. This paper presents a real-time, multimodal, high sample rate lower-limb motion capture framework, based on wireless wearable sensors and machine learning algorithms. *Methods:* Random Forests are used to estimate joint angles from IMU data, and ground reaction force (GRF) is predicted from instrumented insoles, while joint moments are predicted from angles and GRF using deep learning based on the ResNet-16 architecture. *Results:* All three models achieve good accuracy compared to literature, and the predictions are logged at 1 kHz with a minimal delay of 23 ms for 20s worth of input data. *Conclusion:* We were able to achieve decently accurate predictions with minimal delay. *Significance:* The present work fully relies on wearable sensors, covers all five major lower limb joints, and provides multimodal comprehensive estimations of GRF, joint angles, and moments with minimal delay suitable for biofeedback applications.**

## I. Introduction

WALKING is a basic locomotion skill and consists in most of human daily movement; it involves interaction of bony alignment, joint range of motion, neuromuscular activity, and the physics rules of motion, which makes it a relevant activity to be studied in the field of biomechanics, with gait analysis specializing in the assessment of gait and detection of abnormities [1]. Joint angles and moments and ground reaction force are key variables in gait analysis. Joint angles describe the orientation of two articulated segments at a joint and are clinically used to assess joint function as part of gait analysis procedures [2]. Joint moments describe the forces acting on joints and are used in the evaluation of motor function [3], [4], [5], [6], [7] as well as in the design and control of assistive devices such as prostheses [8], [9] and exoskeletons [10], [11], [12], [13]. GRF is important in evaluating ground contact, and is often used along with joint angles to compute joint moments [14].

The traditional golden standard procedure to measure joint angles during motion is to use an optical motion capture system based on cameras and markers placed on the human body to record the movement and then perform an inverse kinematics analysis on the marker trajectories to extract the angles. On the other hand, GRF measurement is generally based on force plates and instrumented treadmills. Joint moments are then calculated using an inverse dynamics analysis from the kinematics results along with the GRF data. Both inverse kinematics and inverse dynamics are performed offline after data collection in a musculoskeletal simulation software such as OpenSim [14] and Anybody [15]. However, optical motion capture systems and force plates are expensive and require a suitable lab space to be installed in, the measurement is limited by the number of available force plates [16], [17], which also impacts the motion of the subject [18], [19], and marker placement as well as the data processing procedures are time-consuming and often requiring manual setup and specific technical skills [20], meaning that this procedure cannot be deployed in real time outside the lab, such as in everyday environments or within wearable assistive devices [1]. While markerless motion capture has recently been proposed as a more portable, less skill- and time-consuming solution [20], [21], commercial systems are still relatively expensive and require the use of powerful computers [22].

This work was supported by the MathWorks-CUED Small Grant Programme 2024 and 2025. J. Mallah is supported by the Cambridge Trust via an Allen, Meek, and Read Cambridge International Scholarship. L. G. Occhipinti acknowledges funding from UK Research Council (EPSRC (grants EP/W024284/1, EP/P027628/1) and from the British Council (UKIERI Contract No. 45371261).

J. Mallah, Yu Zhu, Kailang Xu, and Luigi G. Occhipinti* are with the Electrical Engineering Division, Department of Engineering, University of Cambridge, Cambridge CB3 0FA, United Kingdom (correspondence e-mail: lgo23@cam.ac.uk). Yu Zhu is with the Department of Electronic Systems, Aalborg University, Aalborg 9220, Denmark. Gurvinder Virk is with Endoenergy Systems Ltd., Cambridge, United Kingdom, and with Huzhou Wuxing District Intelligent Robot Innovation Research Institute, China. Shaoping Bai is with the Department of Materials and Production, Aalborg University, Aalborg 9220, Denmark.

All experimental procedures involving human subjects were approved by the Department of Engineering Ethics Committee at the University of Cambridge (Reference No. 639). The study was conducted in accordance with institutional guidelines and the Declaration of Helsinki.

Supplementary material, consisting mostly of additional figures and tables, are available online.

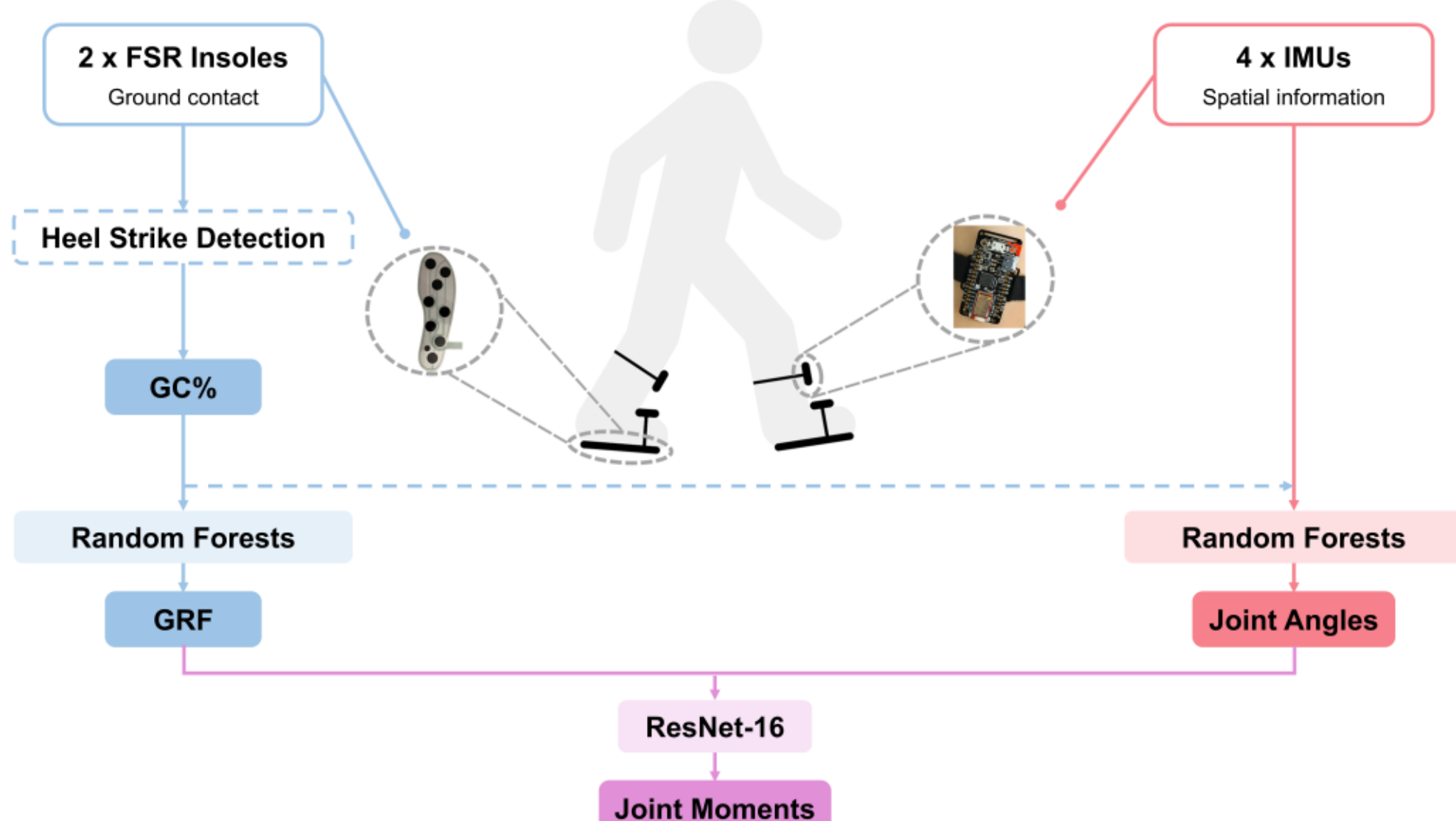

**Fig. 1.** In this paper, Random Forests are used to estimate joint angles from IMU data, and GRF from instrumented insoles, while joint moments are predicted from angles and GRF using deep learning using a model based on the ResNet-16 architecture.

For purposes of joint angle tracking, inertial measurement units (IMUs) are seen as a low-cost alternative for real-time deployment, able to provide segment orientation estimates by integrating linear acceleration, angular velocity, and magnetometer data and using sensor-fusion algorithms such as Kalman filters, complementary filters, and gradient descent algorithms [2], [23], [24]. This process faces several challenges, including sensor drift and difficulties aligning the sensors to body segments and anatomical references [2], [23], [25], but machine learning algorithms can be used [2], [23], [26], [27], [28] and seem to overcome these challenges. Similarly, sensor-instrumented insoles have recently been proposed, in particular those embedded with force sensitive resistors (FSRs) [29], [30], [31], [32], and were shown to be a good wearable alternative to force plates, being unobtrusive, inexpensive, and having a low power consumption, while not affecting gait patterns [2].

This paper presents a real-time, multimodal, high sample rate lower-limb motion capture framework (Fig. 1). The Random Forests algorithm is used to predict ankle angles using IMUs placed on the shank and foot, as well as to predict the vertical ground reaction force (vGRF) from flexible wearable insoles embedded with force myography (FMG) sensors. Weight-normalized vGRF is then added to the feature set of the angle prediction model, allowing for better predictions. Deep learning models based on the ResNet-16 architecture are then used as a replacement for inverse dynamics analyses to estimate ankle joint moments from the ankle angle and vertical GRF. The study is then extended to predict five lower limb joint angles and moments – hip flexion, adduction, and rotation, knee flexion, and ankle flexion.

## II. Methods

### A. Data Collection

8 young healthy subjects were recruited for the experiment (3 females and 5 males, age: 24.50 ± 2.92 years, height: 177.63 ± 8.50 cm, weight: 75.75 ± 11.72 kg). Motion capture data was collected using 16 Vicon Valkyrie VK26 cameras (©Vicon Motion Systems Ltd., Oxford, UK), 41 reflective markers (9 mm, ©B&L Engineering, California, USA), and 3 AMTI ground embedded force plates (2 of 46 × 51 cm, and 1 of 40 × 60 cm, ©Advanced Mechanical Technology, Inc., Watertown USA). Additionally, 4 thermal FLIR cameras (© Teledyne FLIR LLC, Oregon, USA) provided video recording. 9-axis (3D accelerometer, gyroscope, and magnetometer) commercial and open-source EmotiBit IMUs (© Connected Future Labs LLC, Nevada, USA) were worn bilaterally on the shanks and feet, and the data was collected at 25 Hz and transmitted wirelessly to a computer via UDP, but data from all 4 IMUs was only available for four subjects due to sensors disconnecting (1 female and 3 males, age: 23.75 ± 2.05 years, height: 180.75 ± 10.96 cm, weight: 82 ± 10.86 kg). Five subjects (2 females and 3 males, age: 25.20 ± 3.43 years, height: 179.40 ± 7.99 cm, weight: 76 ± 11.66 kg) wore sandals instrumented with insoles (design inspired by [2]) embedded with 8 FSR-based FMG sensors [19] each. All subjects walked in straight lines at their self-selected speeds.

### B. Data Pre-Processing

An OpenSim model is scaled to the dimensions of each subject using static trial data, and the inverse kinematics tool is executed using marker trajectories from the dynamic trials; the gait2392 model is used. The kinematics results are fed into the inverse dynamics tool along with the force plate data to compute the joint moments. Joint angles and moments are up-sampled from 100 Hz to 1 kHz. The IMU data is also up-sampled to 1 kHz using linear interpolation, and bandpass filtered using a fifth-order Butterworth filter with cut-off frequencies of 0.2 and 10 Hz. The marker and IMU data were synchronized using UNIX timestamps and trimmed to the gait cycle duration, which was delimited as follows: for each walking trial, 3 heel strikes are identified from force plate data (20 N threshold), and the fourth heel strike that marks the end of the second stride is determined using the velocity of the heel markers. Hence gait cycle percentages (GC%) corresponding to each datapoint could be generated. The force plate and FSR data were synchronized using UNIX timestamps and trimmed to the gait cycle duration. The vertical component of the ground reaction force was considered and normalized by body weight. FSR data is lowpass filtered at 3 Hz using a fifth-order Butterworth filter. GRF data used in the angle models are normalized by body mass.

### C. Machine Learning Models

Model input data was scaled using a Standard Scaler to have zero mean and unit variance. All models were trained using two methods: (1) intra-subject, performing k-fold cross validation (k=5 for GRF and angles, k=4 for moments) while combining the data from all subjects, shuffling it, and splitting it by gait cycle, and (2) inter-subject, applying *LeaveOneSubjectOut* Cross Validation (LOSOCV).

#### 1) GRF Model

The Random Forest algorithm is used, for its robustness to relatively small datasets due its non-parametric learning method apart from the number of trees [2, 20, 21]. Instead of using time-series machine learning algorithms, GC% is used along with the FSR data for vGRF prediction to add a temporal dimension to the network that is independent of the walking speed, while simplifying the model independently from hyperparameter tuning and significantly reducing the risk of overfitting and training time. The model was trained using Scikit Learn default settings (100 trees).

#### 2) Angle Models

The models were also developed using the Random Forest algorithm, which is preferred over deep learning algorithms such as CNN or LSTM to reduce complexity when low latency is crucial [33], [34], [35], which makes Random Forests the best trade-off in terms of speed, accuracy, interpretability, and simplicity [26], [35], [36]. For the walking trials, and instead of using time-series machine learning algorithms, GC% is used as an input to add a temporal dimension to the network that is independent of the walking speed, and can also be obtained in real-time from insoles or heel contact sensors [37], [38], along with a Boolean flag to indicate the leading foot (on which GC% calculation is based). The main model inputs are 9 channels from each IMU. Models use inputs from either 2 IMUs from the same leg to predict the corresponding ankle angle (unilateral), or 4 IMUs from both legs to predict both ankle angles (bilateral). The study was then extended to predict five lower limb angles: hip flexion, adduction, and rotation, knee flexion, as well as ankle flexion. Several models were developed using different combinations of inputs and outputs as shown in Table I. Model settings were specified in Scikit Learn (200 trees, max_depth=None, min_samples_split=2, min_samples_leaf =1).

TABLE I
DIFFERENT MODEL INPUT AND OUTPUT COMBINATIONS

| Model Name | Inputs | Output Angles |
|---|---|---|
| W1 | 20 (2 IMUs, GC%, flag) | 1: ankle |
| W2 | 21 (2 IMUs, GRF, GC%, flag) | |
| W3 | 40 (4 IMUs, 2 GRF, GC%, flag) | 2: ankles (right, left) |
| W4 | 20 (2 IMUs, GC%, flag) | 5: 3D hip, knee, ankle |
| W5 | 21 (2 IMUs, GRF, GC%, flag) | |
| W6 | 40 (4 IMUs, 2 GRF, GC%, flag) | 10: 3D hips, knees, ankles (right, left) |

#### 3) Moment Models

A deep learning model based on the ResNet-16 architecture and 1D convolutional neural networks (CNN) layers was used. This a time-series model, and similar architectures have been used for similar applications [39], [40], [41]. The model starts with an initial convolutional block including 1D convolution, batch normalization, and ReLU activation layers, to help reduce the time dimension while extracting features. 4 residual blocks follow, allowing for deep feature extraction without vanishing gradients. A global average pooling operation follows, along with a fully connected dense layer and an output layer consisting of 2 neurons, one for each output variable. The network was trained using the ADAM optimizer and mean squared error (MSE) loss over 500 epochs, while implementing early stopping with a patience of 10 epochs and restoring the best weights.

The input ankle joint angles and GRF data were scaled using a Standard Scaler to have zero mean and unit variance. Two other inputs were added: GC%, and a Boolean flag to indicate the leading foot. The input data were passed to the model in windows of 10ms.

In a second instance, we extend our model to predict the five main lower limb joint moments – hip flexion, adduction, and rotation, knee flexion, and ankle flexion – based on the corresponding joint angles and vertical GRF, along with GC% and the lead/lag flag. The same ResNet-16 architecture is deployed and trained using the same intra-subject and inter-subject methods.

#### 4) Performance Evaluation

vGRF and joint moment estimation performance was assessed using the mean absolute error (NMAE) and the mean root mean square error (NRMSE) both normalized by range. Additionally, the Pearson correlation coefficient (r) and the coefficient of determination ($r^2$) were computed for predictions and test data. Angle estimation performance on the test set was assessed using the mean root mean squared error (RMSE).

#### 5) Model Chaining

While joint angles and GRF data used in training and evaluating the moment prediction model were obtained using the normal procedure involving motion capture systems and musculoskeletal models, we want to test the applicability of our model to real-time walking scenarios, using data provided by wearable sensors, i.e. IMUs and FSRs. Therefore, the predictions provided by the angle and GRF models will feed into the ankle moment prediction model.

*6) Real-Time System*

Near real-time motion capture, including ground reaction force data as well as lower limb joint angles and moments during walking of healthy subjects is achieved using 2 IMUs and one FSR-instrumented insole per leg.

The code is developed in Python and run wirelessly on a computer 13th Gen Intel® Core™ i7-13700, 2100 MHz, 16 Cores, 24 Logical Processors. The program collects data from the sensors, and performs timestamp adjustment, up-sampling (IMU), and filtering. Heel strikes are detected from the insole data and used to calculate the gait cycle percentage needed for the predictors (from the previous stride). GRF and angle predictions are then made based on the insole and IMU data, respectively (GRF was not integrated in angle prediction: model W4), and feed into the moment predictor for estimating the joint moments. The extended angle and moment models were used, providing data for the five major lower limb joint angles: hip flexion, adduction, and rotation, knee flexion, and ankle flexion. The angle and moment predictions obtained are rather noisy, therefore a low pass filter was applied to clean it. All the data is saved in log files: insole raw, timestamp-adjusted, and filtered; IMU raw, up-sampled, and filtered; GRF predictions; raw and filtered angle and moment predictions.

The program was tested on a healthy subject (age: 21 years, height: 178 cm, weight: 59 kg), using a 20s trial time for data collection.

To evaluate the quality of the predictions, we calculate Pearson's r between the mean profile of the initial dataset (which was used for training and offline testing) and real-time (RT) predictions

## III. Results

### A. GRF Model

The vGRF model achieves an NRMSE of 5.09 ± 0.61% in intra-subject mode, and 8.36 ± 0.91% in inter-subject mode.

### B. Angle Models

Table II summarizes the results obtained with the different models, among which W6 is the best performing.

### C. Moment Models

For the ankle-only model, the intra-subject method achieved an NRMSE of 4.88 ± 0.83% with a correlation coefficient of 0.97 ± 0.01, while the inter-subject method resulted in an NRMSE of 7.72 ± 1.60% with a correlation coefficient of 0.93 ± 0.03.

Extending to the 5-joint models, the intra-subject method achieved a minimum NRMSE of 1.63 ± 0.03% for hip rotation moment prediction and a maximum NRMSE of 2.96 ± 0.03% for hip adduction; the lowest correlation coefficient of 0.97 ± 0.00 was obtained for hip adduction, whereas the knee joint had the best correlation coefficient of 0.9952 ± 0.00. As for the inter-subject method, a minimum NRMSE of 6.59 ± 0.64% for hip flexion moment prediction and a maximum NRMSE of 10.50 ± 1.70 for the ankle joint were obtained, along with a minimum correlation coefficient of 0.98 ± 0.00 for the joint, and a maximum correlation coefficient of 0.90 ± 0.01 for the joint. The full results are shown in Table III.

### D. Real-Time System

The GRF, angle, and moment predictor blocks finish 5, 14, and 23 ms after the end of the 20s data collection window. Figs. 2–4 show the real time predictions versus the average curves obtained during the initial offline data collection across all subjects, along with the standard deviation of the initial data and offline predictions (from testing in sections 3.1 to 3.3) combined. Table IV shows the r and $r^2$ values for all prediction stages.

## IV. Discussion

### A. GRF Model

In this work, we standardized the FSR sensor data and applied weight normalization to GRF, as FSRs are unable to measure kinetic parameters, i.e. weight-induced changes in GRF, but are capable of capturing the relative distribution of plantar pressure. Hence, the values predicted by the network have to be scaled by the subject's weight to yield the true GRF [31].

Machine learning techniques are often used to estimate GRF from insole data. Oubre *et al.* [31] used the Random Forest algorithm to predict triaxial GRF and obtained a total NRMSE of 4.9% and a per-subject NRMSE of 7.7% (BW) for the

TABLE II
Summary Table Compiling the RMSE Results from all the Angle Models

| Model (I/O) | Mode | Hip Flexion | Hip Adduction | Hip Rotation | Knee Flexion | Ankle Flexion |
|---|---|---|---|---|---|---|
| W1 (20/1) | Intra | | | | | 5.45 ± 0.22° |
| | Inter | | | | | 9.06 ± 1.80° |
| W2 (21/1) | Intra | | | | | 5.11 ± 0.18° |
| | Inter | | | | | 9.08 ± 2.14° |
| W3 (40/2) | Intra | | | | | 4.41 ± 0.56° |
| | Inter | | | | | 8.71 ± 2.21° |
| W4 (20/5) | Intra | 6.40 ± 0.91° | 3.23 ± 0.12° | 5.30 ± 0.25° | 7.48 ± 1.21° | 5.45 ± 0.30° |
| | Inter | 9.14 ± 2.96° | 4.81 ± 0.96° | 7.03 ± 1.76° | 11.00 ± 3.69° | 9.06 ± 2.06° |
| W5 (21/5) | Intra | 5.61 ± 0.38° | 3.15 ± 0.10° | 5.21 ± 0.12° | 6.09 ± 0.88° | 5.08 ± 0.18° |
| | Inter | 7.48 ± 1.53° | 4.69 ± 0.99° | 7.32 ± 1.60° | 9.39 ± 2.46° | 8.66 ± 2.04° |
| W6 (40/10) | Intra | 4.80 ± 0.70° | 2.60 ± 0.14° | 4.57 ± 0.38° | 5.19 ± 1.56° | 4.16 ± 0.59° |
| | Inter | 8.15 ± 1.84° | 4.97 ± 1.15° | 7.67 ± 2.16° | 10.07 ± 3.08° | 8.39 ± 2.44° |

TABLE III

EVALUATION METRICS OF THE DEEP LEARNING MODELS IN INTRA- AND INTER-SUBJECT MODES

| Method | Joint | NRMSE (%) | NMAE (%) | $R^2$ | r |
|---|---|---|---|---|---|
| Intra-subject | Hip Flexion | 2.48 ± 0.06 | 1.85 ± 0.03 | 0.98 ± 0.00 | 0.99 ± 0.00 |
| | Hip Adduction | 2.96 ± 0.03 | 2.27 ± 0.01 | 0.94 ± 0.00 | 0.97 ± 0.00 |
| | Hip Rotation | 1.63 ± 0.03 | 1.20 ± 0.02 | 0.98 ± 0.00 | 0.99 ± 0.00 |
| | Knee Flexion | 1.87 ± 0.12 | 1.04 ± 0.03 | 0.99 ± 0.00 | 1.00 ± 0.00 |
| | Ankle Flexion | 1.68 ± 0.08 | 1.22 ± 0.05 | 0.98 ± 0.00 | 0.99 ± 0.00 |
| Inter-subject | Hip Flexion | 6.59 ± 0.64 | 4.85 ± 0.55 | 0.78 ± 0.04 | 0.90 ± 0.01 |
| | Hip Adduction | 9.90 ± 0.65 | 6.87 ± 0.49 | 0.75 ± 0.05 | 0.87 ± 0.03 |
| | Hip Rotation | 8.23 ± 1.35 | 6.11 ± 1.07 | 0.67 ± 0.08 | 0.84 ± 0.04 |
| | Knee Flexion | 8.17 ± 1.29 | 5.82 ± 0.86 | 0.67 ± 0.10 | 0.87 ± 0.04 |
| | Ankle Flexion | 10.50 ± 1.70 | 5.57 ± 1.38 | 0.74 ± 0.08 | 0.88 ± 0.03 |

vertical component, with an $r^2$ of 0.91. Zhang *et al.* [29] used a deep dual-stream cross attention model to estimate triaxial GRF, and obtained an NRMSE of 4.16% (BW) without using cross validation or leaving subjects out. Hajizadeh *et al.* [30] estimated 3D GRF using LSTM networks, resulting in an NRMSE of 5.5% (BW). Choi *et al.* [32] used 3 different network architectures for GRF prediction, based on Artificial Neural Networks (ANN), CNN, and a combination of CNN and Long Short-Term Memory (LSTM) networks; however, they did not apply body weight normalization to the input data and reported the RMSE in N which makes it difficult to compare results. Assuming GRF can reach up to 1.2 times body weight [42], with the heaviest subject having a mass of 79 kg, their best NRMSE can therefore be approximated to 16.77%. Hence, our models achieve comparable results to the literature.

### B. *Angle Models*

Our purpose was to estimate lower limb joint angles based on data from 9-axis IMUs placed on the shanks and feet using machine learning based on the Random Forest algorithm. Looking at the summary Table II, among the ankle-only models, the lowest RMSEs of 4.41 ± 0.56° and 8.71 ± 2.21° in intra- and inter-subject modes, respectively, were obtained with the bilateral 4 IMUs + 2 vGRF model (W3). Expanding the 5-joint models, the same trend holds, as W6 performs the best (4.16 ± 0.59° and 8.39 ± 2.44°). W5 performs better than W4, and despite W2 having a similar performance to W1 in inter-subject mode (harder generalization which can be attributed mostly to inter-subject variability over more input variables), it performs better in intra-subject mode, which reflects the benefit of adding vGRF to the feature set on the quality of the predictions. The expansion to bilateral models seems beneficial, as W3 and W6 perform better than W2 and W5, respectively, and S2 best (5.79 ± 0.17° and 11.20 ± 4.01°) performs better than S1. The addition of GRF and bilateral expansion in intra-subject mode gradually improved the performance of ankle-only models (W3 better than W2 better than W1) in the same way as the 5-joint models (W6 better than W5 better than W4), which seems logical as adding more relevant features provides more contextual information and normally improves the performance of machine learning models.

When expanding to the 5-output model, it is interesting how IMUs placed on the shanks and feet can contribute to the prediction of not only neighboring distal ankle and knee angles, but also proximal 3D hip angles, showing that all joint angles are closely related, which was taken by [28] as far as predicting hip, knee, and ankle angles with only one IMU placed on the shank.

While lower-limb joint angles have been obtained from IMUs without using machine learning, such as in [43], [44], or using different inputs such as surface EMG [45], machine learning has been frequently used to predict angles from IMU data. Ackland *et al.* [23] used accelerometer and gyroscope data from IMUs placed on the pelvis, thighs, shanks, and feet to estimate 3D angles at hip, knee, and ankle with a generative adversarial network (GAN), resulting in an RMSE of 1.80 – 5.30° for ankle flexion, down to 0.6° for hip rotation (0.1° for knee abduction) and up to 7.30° for knee flexion. Similarly,

TABLE IV

PEARSON CORRELATION COEFFICIENT (R) AND THE COEFFICIENT OF DETERMINATION ($R^2$) BETWEEN REAL TIME PREDICTIONS AND INITIAL OFFLINE DATA COLLECTION AVERAGES ACROSS ALL SUBJECTS

| Model Name | Motion | r | $r^2$ |
|---|---|---|---|
| GRF | | 0.98 | 0.96 |
| Angles | Hip Adduction | 0.99 | 0.97 |
| | Hip Rotation | 0.93 | 0.73 |
| | Hip Flexion | 0.88 | 0.62 |
| | Knee Flexion | 0.99 | 0.98 |
| | Ankle Flexion | 0.96 | 0.87 |
| Moments | Hip Adduction | 0.94 | 0.79 |
| | Hip Rotation | 0.96 | 0.91 |
| | Hip Flexion | 0.94 | 0.85 |
| | Knee Flexion | 0.86 | 0.62 |
| | Ankle Flexion | 0.96 | 0.92 |

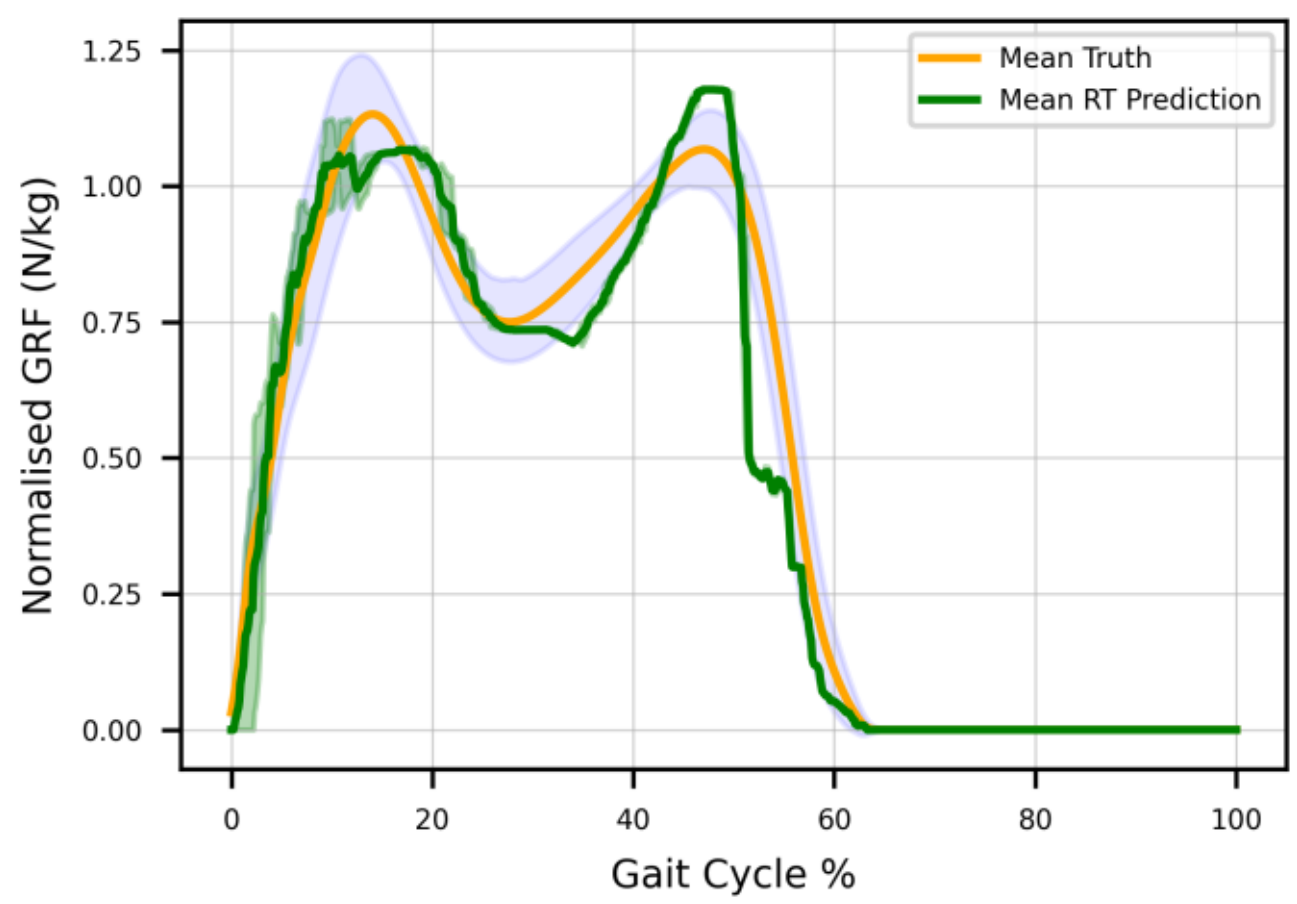

**Fig. 2.** Real-Time GRF versus initial ground truth data and offline predictions.

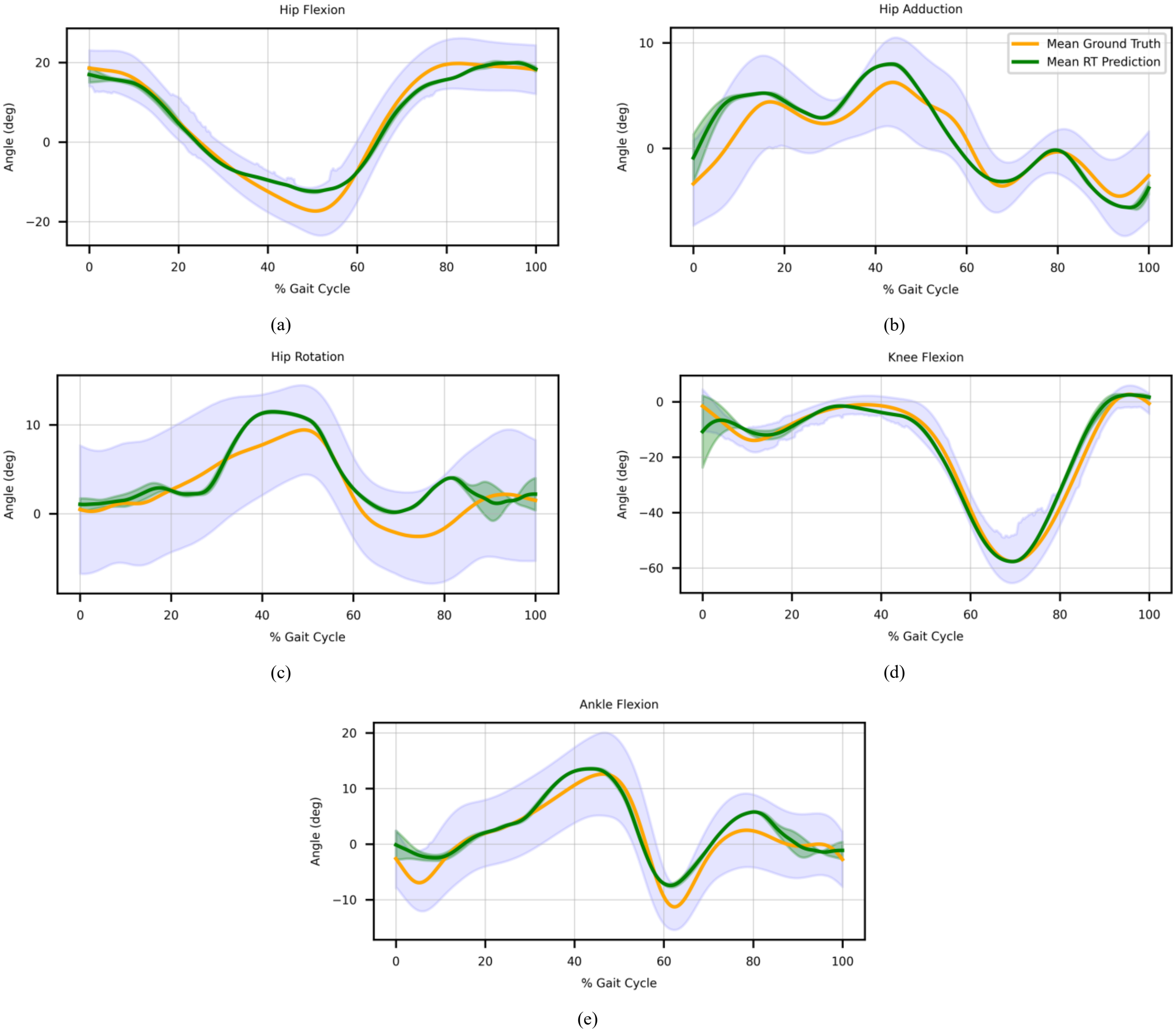

**Fig. 3.** Ensemble averages of the real-time filtered joint angles (green) vs initial dataset (orange). The green shading represents the standard deviation of the real-time filtered data. The blue shading corresponds to the standard deviation of the initial data and the offline predictions combined.

Senanayake *et al.* [2] deployed a GAN on accelerometer and gyroscope data from IMUs placed on the thighs, shanks, and feet to predict 3D ankle angles, achieving an RMSE of 3.60 – 5.90° for ankle flexion (down to 1.70° for ankle inversion). Hollinger *et al.* [26] also used accelerometer and gyroscope data from IMUs placed on the torso, thighs, shanks, and feet, but with Random Forests to predict hip, knee, and ankle flexion angles during isolated joint movements. They obtained an RMSE of 5.35 ± 4.28° for the ankle with 2 IMUs and 5.55 ± 4.10° with 4 IMUs, while the knee RMSE reached 20.71 ± 12.41° with 4 IMUs. Hur *et al.* [27] used data from 9-axis IMUs placed at the torso, pelvis, thighs, shanks, and feet to predict hip, knee, and ankle flexion angles with convolutional neural network (CNN) and long-short term memory (LSTM) architectures, achieving an RMSE of 2.45 – 7.98° at the ankle with 2 IMUs (shank and foot), up to 8.30° with different IMU combinations/training methods, and up to 17.06° for knee angles. On the other hand, Sung *et al.* [28] used only 1 IMU placed on the shank to predict ankle, knee, and hip flexion angles using an LSTM network, achieving an RMSE of 0.42 – 5.15° at the ankle with an $r^2$ of 0.62 – 0.96, and up to an RMSE of 8.14° at the knee. Therefore, this work achieves comparable results to the literature, while predicting relatively multiple lower limb angles with a simple low latency algorithm suitable for real-time implementation.

### *C. Moment Models*

Our goal was to estimate ankle moments based on ankle angles and GRF using machine learning techniques. A time-series model based on the ResNet-16 architecture is introduced, achieving NRMSEs as low as 4.88%. The model was then expanded to the prediction of five lower limb moments – hip flexion, adduction, and rotation, knee flexion, and ankle flexion – using the corresponding joint angles and GRF as inputs for a similar ResNet-16 model, achieving even lower NRMSEs in the range of 1.63 – 2.96%. An expected drop in performance occurs when evaluating the model on unseen subjects in inter-subject mode, but the results remain acceptable (Table III), proving the robustness of our model.

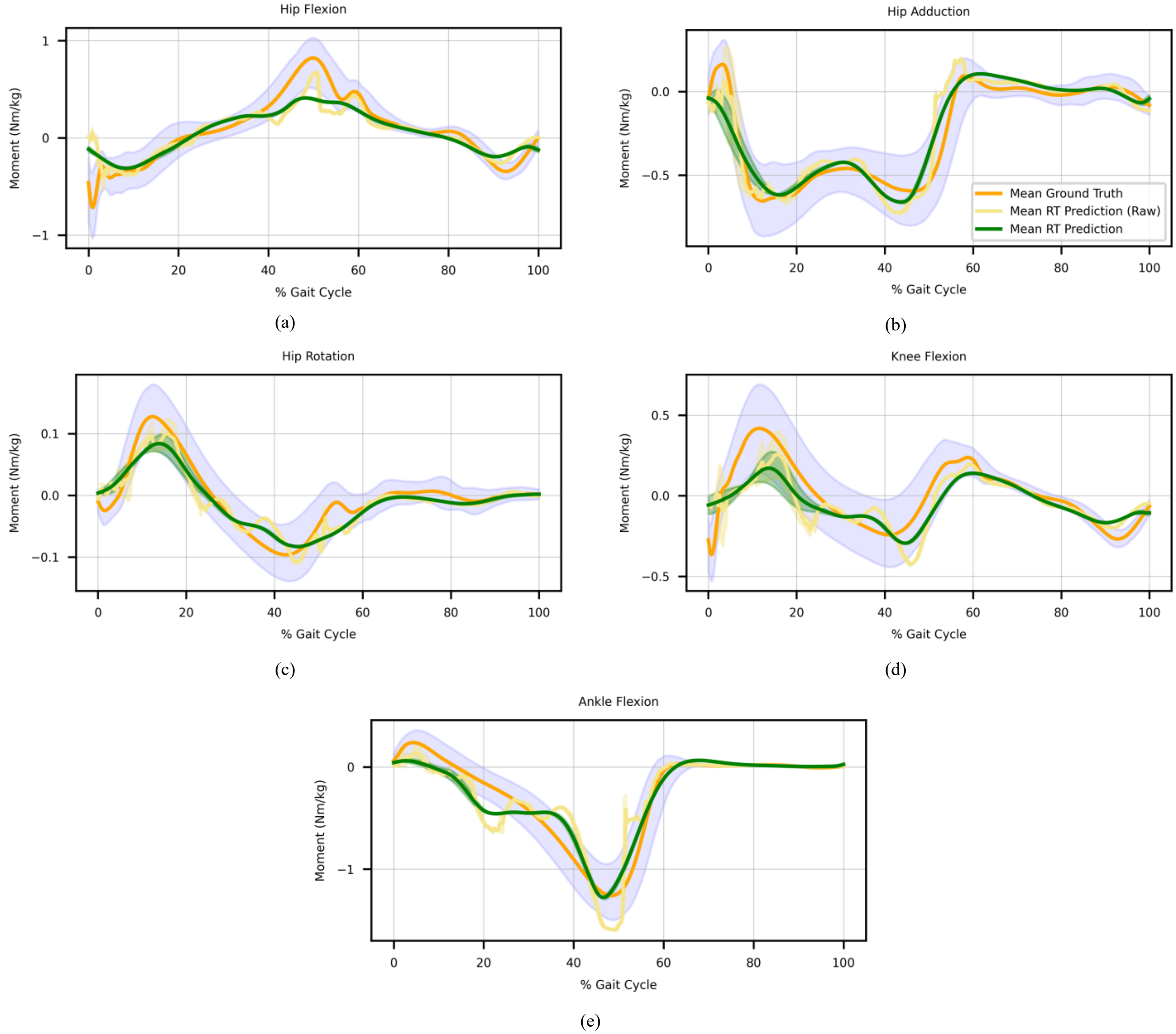

**Fig. 4.** Ensemble averages of the real-time unfiltered (purple) and filtered (green) joint moments vs initial dataset (orange).

While completely different inputs have been used for joint moment prediction in the literature, such as IMUs [46], a combination of electromyography (EMG), IMUs, and electrogoniometers [47] or EMG and muscle synergies [48], joint angles and GRF were used as well. Perrone *et al.* [49] estimated hip moments from knee flexion angles and GRF using a Long-Short Term Memory (LSTM) architecture, yielding an NRMSE of 9.62% and 15.55% NMAE. Giarmatzis *et al.* [50] predicted 3D medial and lateral knee contact forces from 13 angles spanning the human body and 3D GRF using Artificial Neural Networks (ANN) and Support Vector Regression (SVR), achieving an NRMSE of 0.67 – 5.39%. Ozates *et al.* [41] expanded to 15 angles as inputs, and estimated ankle (flexion), knee (flexion), and hip (abduction and flexion) moments with 1D CNNs, yielding an NRMSE of 8.58 ± 3.87% at the ankle in healthy subjects, compared to 14.78 ± 7.17% in participants with cerebral palsy, up to 12.55 ± 5.08% and 18.02 ± 9.14%, respectively, at the knee. Mundt *et al.* [19] had slightly different settings, this time predicting 3D GRF along with 3D moments, given ankle, knee, and hip joint angles, using Feed-Forward (FF) neural networks and LSTM, and obtained an NRMSE of 8.69 – 15%. Other studies also estimated joint moments but using slightly different input combinations. For example, Xiong *et al.* [13] used five joint angles (3D hip, knee flexion, ankle flexion) and 10 EMG channels to predict 4 joint moments (2D hip, knee flexion, ankle flexion) with an ANN, resulting in an NRMSE of 6.70% at the ankle and 7.89% at the hip. Ardestani *et al.* [51], on the other hand, combined 2D GRF with 8 EMG channels to predict 3D hip, knee flexion, and 2D ankle angles using an FF ANN and Wavelet Neural Networks (WNN), resulting in ankle flexion NRMSE in the range of 5 – 12%, and up to < 16% for hip adduction. Therefore, in comparison, our models achieve remarkable results with a minimal number of inputs.

### D. *Real-Time System*

The objective was to achieve near real-time lower-limb motion capture, including joint angles, moments, and GRF, aiming to substitute gold-standard devices and musculoskeletal processing with a few wearable sensors and machine learning algorithms. This target was reached as the new method executes swiftly and achieves very high correlation (Pearson's r >= 0.88) with the expected values (Table IV), and acceptable $r^2$ values (>= 0.79), except for hip rotation (0.73) and flexion (0.62) angles which tend to be relatively overestimated and knee flexion moment (0.62) which is rather underestimated, by reference to Figs. 3 and 4, but remaining mostly within one standard deviation.

Previous related work includes [52] which determines the full 3D skeletal pose in joint angles using a single RGB camera and an algorithm combining CNN and fully-connected NN, achieving a > 30 fps speed with a 32.5 ms delay, [53] which estimates 3D pose using 4 cameras and up to 17 IMUs at 40 fps and overall latency of 230 ms, using a solver that optimizes the pose cost function, and [54] which determines the 3d full-body pose using acceleration and rotation data from 3 IMUs along with a combination of transformer encoder, bidirectional LSTM and multilayer perceptrons, at 60 fps with a 166ms delay. These three studies focused on full body pose, and while [52] predicted joint angles specifically, [53] adopted 3D angle-axis vectors (i.e. the axis of rotation multiplied by the angle of rotation in radians), and [54] predicted the global rotation of the pelvis and the local rotation of other joints relative to the parent joints, represented by 6-D rotation. While [52] and [53] used cameras, [55] and [56] still used standard motion capture devices, but focused on achieving a real-time data processing. [55] combined a Kalman filter, a multibody formulation, and an optimization algorithm to estimate skeletal kinematics (position, velocity, and acceleration), joint torques, and muscle efforts, respectively, with a 10 ms delay (0.8 ms inference time), and [56] which managed to use OpenSim in real time, achieving up to 2,000 fps with 31.5 ms delay. Closer to our work and [54], i.e. using wearable sensors only, [57] predicted knee angles and hip moments with an inference time of around 3 ms, using two soft stretchable capacitive sensors integrated into a knee pad. Other studies claim to have achieved real-time estimation of joint angles or moments but without providing sufficient information on the real-time implementation itself (e.g. hardware and setup) and latency, such as [50] which predicts knee moments by fusing optical motion capture and musculoskeletal modelling-derived kinematic and force variables, and [2], [23] which estimate three lower-limb joint angles using up to 3 IMUs.

Therefore, the present work stands out by (1) fully relying on wearable sensors without requiring any cameras or gold-standard motion capture systems, (2) providing estimations at 1000 fps with 23 ms delay per 20 s worth of input data, (3) covering all five major lower limb joints, and (4) providing multimodal comprehensive estimations of GRF, joint angles, and moments. Limitations, however, include the limited size of the dataset used for training, covering only lower-limb joints, with applications restricted to walking, i.e. the movement on which the models were trained [55].

## V. Conclusion

In this article, we presented a real-time, multimodal, high sample rate lower-limb motion capture framework, based on wireless wearable sensors and machine learning algorithms. The developed models achieve good accuracy compared to literature, providing predictions at 1 kHz with a minimal delay of 23 ms for 20s worth of input data. Future work can consider developing separate predictors by movement type and combining them with a motion classifier that picks the model suitable for the current detected activity. Moreover, another possible application is similar to [57], in the control of a lower-limb exoskeleton; however, the present models will need to be validated and possibly retrained for the assisted scenario as kinematic/kinetic changes can be introduced by the exoskeleton.